\documentclass[twocolumn,showpacs,preprintnumbers,amsmath,amssymb,10pt,pra,aps]{revtex4-1}
\usepackage{}

\usepackage[version=3]{mhchem} 
\usepackage{graphicx}
\usepackage{graphics}
\usepackage{epsfig}

\usepackage{epstopdf}
\usepackage{amssymb}
\usepackage{amsmath}
\usepackage{float}
\usepackage{ulem} 

\usepackage[british]{babel}
\emergencystretch=\maxdimen
\begin{document}

\title{Switchable imbibition in nanoporous gold}

\author{Yahui Xue$^{1,2}$, J\"urgen Markmann$^{1,3}$, Huiling Duan$^{2,}$}
\email[Corresponding authors, Email: ]{hlduan@pku.edu.cn}

\author{J\"org Weissm\"uller$^{1,3}$, Patrick Huber$^{3,}$}
\email[Corresponding authors, Email: ]{patrick.huber@tuhh.de}

\affiliation{1 Institute of Materials Research, Helmholtz-Zentrum Geesthacht, D-21502 Geesthacht, Germany}

\affiliation{2 State Key Laboratory for Turbulence and Complex Systems, CAPT, Department of Mechanics and Engineering Science, College of Engineering, Peking University, 100871 Beijing, China}

\affiliation{3  Institute of Materials Physics and Technology, Hamburg University of Technology, D-21073 Hamburg, Germany}

\begin{abstract}
Spontaneous imbibition enables the elegant propelling of nano-flows because of the dominance of capillarity at small length scales. The imbibition kinetics are, however, solely determined by the static geometry of the porous host, the capillarity, and the fluidity of the imbibed liquid. This makes active control particularly challenging. Here, we show for aqueous electrolyte imbibition in nanoporous gold that the fluid flow can be reversibly switched on and off through electric potential control of the solid-liquid interfacial tension, i.e. we can accelerate the imbibition front, stop it, and have it proceed at will. Simultaneous measurements of the mass flux and the electrical current allow us to document simple scaling laws for the imbibition kinetics, and to explore the charge flow dynamics in the metallic nanopores. Our findings demonstrate that the high electric conductivity along with the pathways for ionic and/or fluid transport render nanoporous elemental gold a versatile, accurately controllable electro-capillary pump and flow sensor for minute amounts of liquids with exceptionally low operating voltages.
\end{abstract}

\maketitle




Nanofluidic transport exhibits novel interesting properties, spurring different applications in the analysis of the (bio) chemical and physical properties of small numbers of molecules \cite{Stone2004, Squires2005, Dittrich2006, Whitby2007, Schoch2008, Simon2008, Piruska2010, Kirby2010, Tabeling2014}. One issue of interest is the spontaneous imbibition of fluids in bodies with nanoscale dimensions. Given the dominance of capillary forces at small length scales, it provides an elegant and effective way to propel nanoflows and is frequently employed for the synthesis of novel hybrid materials \cite{Yuan2008, Lang2011, Wang2013} or used in the functionality of nano-devices \cite{Schoch2008, Piruska2010}. Moreover, it can be exploited to explore the rheology of liquids in spatial confinement \cite{Shin2007, Gruener2011} or characteristics of the imbibition geometry on the nanometer scale \cite{Elizalde2014}.

Even for nanoscale pores, theory and experiment confirm the classic Lucas-Washburn law, which predicts a square-root of time imbibition kinetics \cite{Gelb2002,Dimitrov2007, Tas2004,Han2006, Huber2007, Gruener2009, Gruener2011, Gruener2012}. Additionally imbibition dynamics are determined by the geometry of the porous host, the fluid-wall interaction, the fluidity and capillarity of the liquid imbibed. Those properties are static or hardly externally changeable during the transport process, rendering a flexible, active control of the fluid flow, as it is desirable in view of possible applications of imbibition in nano- or microfluidic systems \cite{Whitby2007,Wheeler2008}, very challenging.

Here, we explore a strategy for achieving control over the imbibition kinetics. The distinguishing aspect of our approach is the use of a metallic nanoporous host structure. This allows us to tune the solid-liquid interfacial tension by a potential applied directly to the porous metal. In this way, the driving force for imbibition can be manipulated and -- as we show -- the fluid flow can be switched on and off at will by small voltages.

Macroscopic bodies of nanoporous metal, and specifically nanoporous gold (NPG) are readily fabricated by the controlled electrochemical corrosion (``dealloying") of Ag-Au alloy \cite{LiSieradzki1992,Erlebacher2001,Parida2006,Hodge2007,Jin2011}. Porous metal made in this way exhibits a bicontinuous microstructure, consisting of a uniform network of nanoscale ligaments which is interpenetrated by a geometrically similar open pore space with a pore volume fraction of typically $\sim$70$\%$. The ligament- and pore size is uniform and can be tailored from 5 nm to microns. Reconstruction of the NPG structure by electron tomography confirms the connectivity of the nanopores \cite{Roesner2007,Fujita2008}.
The high electric conductivity along with the pathways for ionic and/or fluid transport in the pore space provides various opportunities for creating novel functional materials in which external strain, electric resistance, or mechanical strength are controlled through electric or chemical signals \cite{KramerNanoLetters2004,BienerNatureMater2009,Wahl2010,Jin2011,Detsi2012,Weissmueller2010}. The efficient transport and storage of charge has also prompted studies of the material for application as a supercapacitor \cite{Lang2011}.

Here, we show that nanoporous gold also presents novel opportunities for investigating and controlling nanofluidic transport. We start out by verifying that spontaneous imbibition of aqueous electrolyte in NPG follows the classic Lucas-Washburn law. We then demonstrate that electrocapillary effects enable the active control of capillary flow. The spontaneous imbibition of aqueous electrolytes into NPG is accelerated when a potential is applied. Furthermore, when the driving force for imbibition is diminished by letting the aqueous electrolyte be imbibed by cyclohexane-saturated NPG, the fluid flow can get switched on and off by the potential.


\noindent{\bf Results}



\begin{figure*}
  \includegraphics[width=18cm]{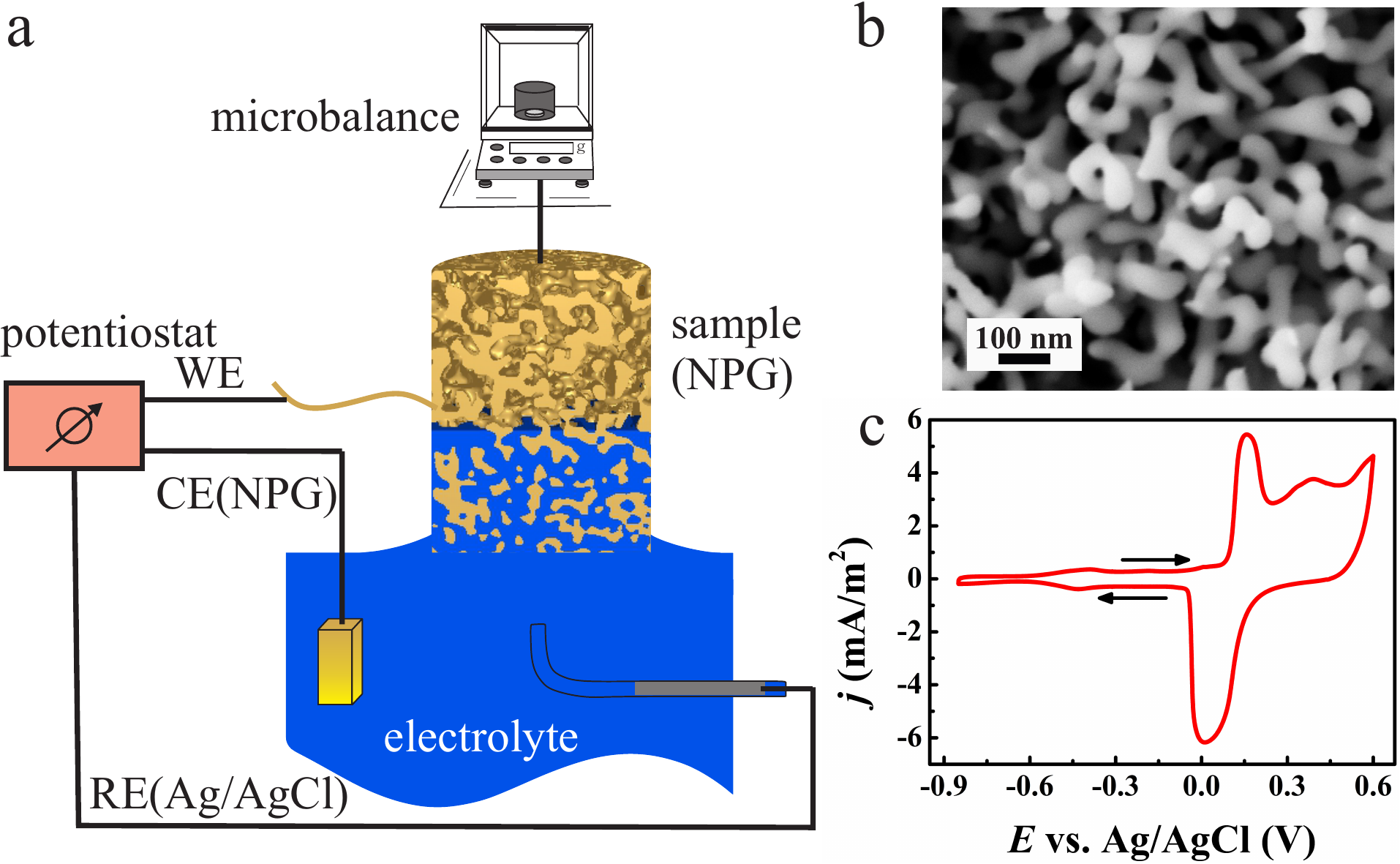}\\
  \caption{{\bf Experimental setup, electron micrograph, and cyclic voltammogram of nanoporous gold.} {\bf a}, Schematics of the experimental setup for imbibition of aqueous electrolytes into nanoporous gold (NPG) under controlled electrode potential. WE, working electrode; RE, reference electrode; CE, counter electrode.  {\bf b}, Scanning electron micrograph showing the microstructure of NPG. Scale bar: 100 nm. {\bf c}, Cyclic voltammogram of current density, $j$, versus potential $E$ in 1 M KOH solution at the potential scan rate of 1 mV/s. The reference electrode is Ag/AgCl in 3M KCl.}
  \label{fig_setup}
\end{figure*}

\noindent{\bf Introductory remarks.} All experiments were carried out at controlled ambient temperature, 22$^\circ$C, using cylindrical NPG samples, 1.08 mm in diameter and 10 mm in length, of uniform ligament size $d$ = 45 nm as determined from an analysis of electron micrographs (see Fig. \ref{fig_setup}b). The synthesis by electrochemical dealloying of a Ag-Au solid solution master alloy made by arc melting and wire drawing is described in Methods. Volumetric nitrogen sorption isotherms were recorded from the samples at a temperature of 77 K. Their analysis by the Barrett-Joyner-Halenda (BJH)-method for capillary condensation in pore space \cite{Lowell2004} indicated a width of the ligament diameter distribution of 25\%. Figure \ref{fig_setup}a displays the experimental setup for the imbibition test under potential control. The sample is suspended over the liquid reservoir from a balance. Spontaneous imbibition starts when liquid and sample make contact as the liquid level is raised by filling the reservoir through a syringe. The electrolyte is 1 M KOH solution. The cyclic voltammogram (CV) of Fig. \ref{fig_setup}c is consistent with published data for Au in alkaline solutions \cite{Lipkowski1999}, testifying in particular to pronounced specific adsorption of $\rm OH^-$ at positive values of the electrode potential, $E$, and a wide region of predominantly capacitive charging at more negative $E$ .

It has been demonstrated that the variation of mass, $m$, with time, $t$, during the spontaneous imbibition of water into hydrophilic nanopores follows the simple scaling law results from Lucas and Washburn  \cite{Gruener2011},
\begin{equation}
    \centering
  m(t)=C \sqrt{t}
  \label{eqn_LW}
\end{equation}
where $C$ represents the imbibition coefficient,
\begin{equation}
    \centering
C=\rho A \phi_{0}
\sqrt{\frac{r_0 \sigma_\mathrm{LV} \cos \theta}{2 \tau \eta}} \,,
  \label{eqn_LW_coefficient}
\end{equation}
with the symbols $\rho$ - density of the fluid, $\eta$ - dynamic viscosity, $\phi_0$ - pore volume fraction, $r_0$ - effective pore radius, $\sigma_\mathrm{LV}$ -
liquid-vapor
surface tension, $\tau$ - tortuosity and $A$ - sample cross-sectional area. The tortuosity describes the connectivity and meandering of the pores \cite{Gruener2012}. Straight channels parallel to the flow direction give $\tau=1$. The meandering of the pores inside porous materials increases the length of the flow path and hence increases $\tau$. For isotropically distributed pores, $\tau$ = 3 is generally expected \cite{Bear1972}, since only one third of the pores contribute to the flow under the pressure gradient. A value of $\tau<3$ indicates relatively good connectivity and low meandering of the pores.


\noindent{\bf Imbibition under open-circuit conditions} We started the imbibition experiment under open circuit conditions. Gold surfaces are hydrophilic when exposed in atmosphere, with a gold-water contact angle $\theta \sim 70^\circ$ \cite{Yokomaku2008,Abdelsalam2005}. Therefore, once the electrolyte touches the NPG sample , it is spontaneously imbibed. A jump of mass at the moment of contact can be attributed to the formation of outside menisci at the liquid-solid-air triple line~\cite{Gruener2009}. Since we are only interested in the capillary flow inside the nanopores, the initial jump has been subtracted from all data. Residual deviation from pure imbibition behaviour can be linked to the evolution of the outside meniscus during the first tens of seconds. Furthermore, the presence of precursor films spreading ahead of the invasion front of nanopores may also contribute to an initial transient behavior ~\cite{Chibbaro2008, Engel2010}.

\begin{figure*}
  \includegraphics[width=8.5cm]{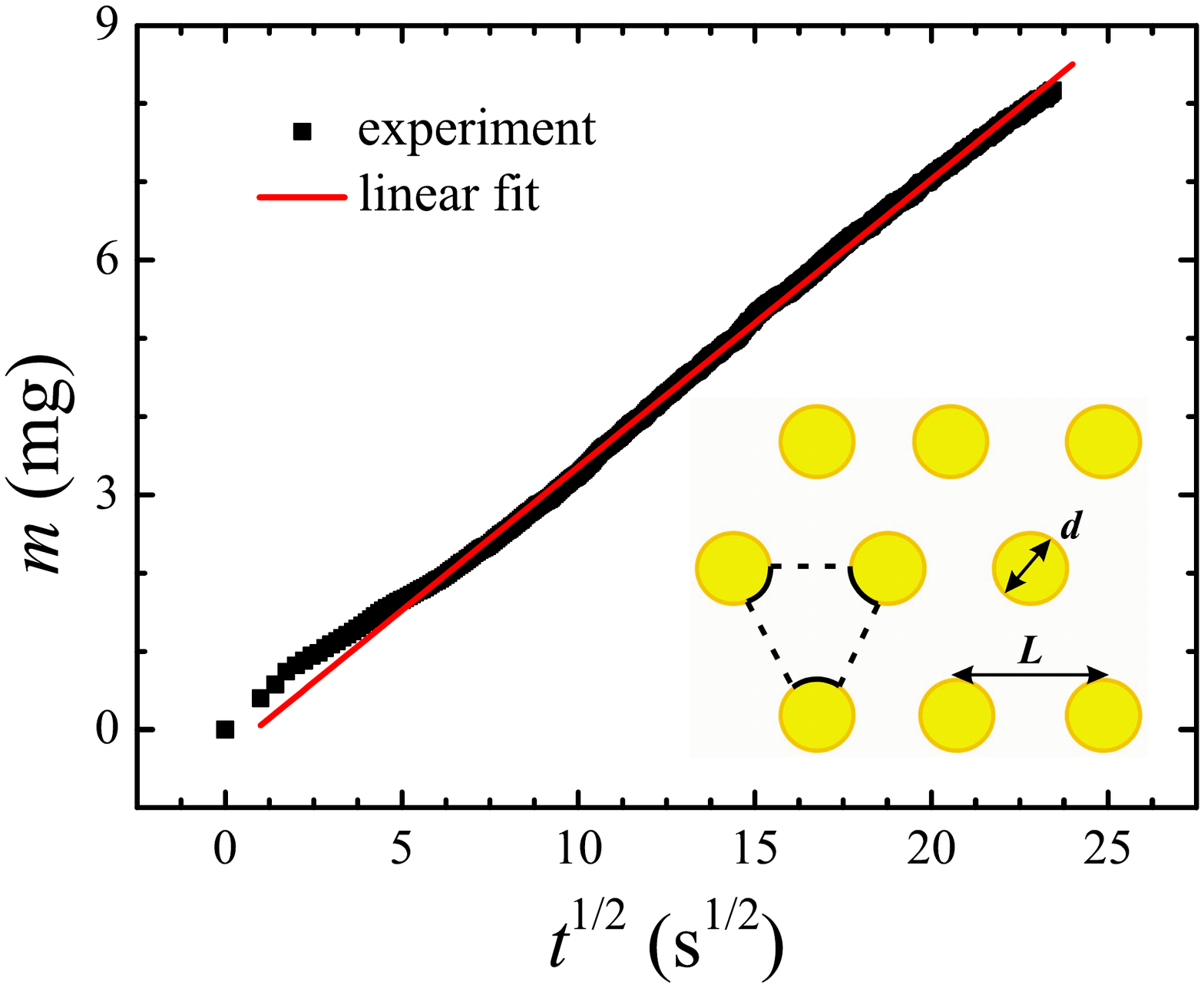}\\
  \caption{{\bf Spontaneous Imbibition.} The graph shows imbibed mass, $m$, versus square-root of time, $t$. The solid line is a linear fit according to equation (\ref{eqn_LW}), Inset: schematics of ligaments in a hexagonal array with a spacing of $L$. The closed curves show the void space and the solid lines represent the arcs.}
  \label{OCP}
\end{figure*}

Figure~\ref{OCP} presents a  plot of the imbibed mass,  $m$, versus the square root of the time, $t^{1/2}$. While the data at $t < 25$ is affected by the meniscus formation, the graph at longer times is linear. This confirms that our data agree with the Lucas-Washburn law. The slope of the linear fit indicates $C =  0.36 \pm 0.01 {\rm mg \cdot s}^{-1/2}$. According to equation (\ref{eqn_LW_coefficient}), $C$ depends on the liquid properties, the structure size and topology of the porous host material, and on the contact angle.

With an eye on estimating the tortuosity based on the experimental $C$ value, we inspect the effective pore radius of the NPG microstructure in Fig. \ref{fig_setup}b. Approximating the ligaments as cylinders of diameter $d$, one may estimate $r_0$ from the analysis of wetting in pillar arrays in ref. \citenum{Lobaton2007}. Here, $ r_0$ is matched to the minimum surface suspended between cylindrical pillars in a hexagonal array, see the inset in Fig. \ref{OCP}. This leads to \cite{Lobaton2007}
\begin{equation}
    \centering
  r_0 = \frac{\phi_0}{2(1-\phi_0)}d.
  \label{eqn_CapRadi}
\end{equation}

Note that previous studies found that, for volume of NPG prepared under the present conditions, the pore volume fraction to agree closely with the less noble metal atom fraction, implying here $\phi_0 = 0.75$ \cite{Jin2009,Wang2013} and, by equation (\ref{eqn_CapRadi}), $r_0=68$ nm. With that result, and using the experimental $C$ value, the density ($\rho$ = 1.046 g/ml~\cite{Akerlof1941}), surface tension ($\sigma_\mathrm{LV}$ = 73.9 mN/m~\cite{Weissenborn1996}) and viscosity ( $\eta$ $\sim$ 1.054 cP) values for 1 M KOH aqueous solution, equation (\ref{eqn_LW_coefficient}) leads to $\tau = 3.2 \pm 0.2$. This result agrees closely with an experimental result for the tortuosity of nanoporous Vycor glass that shows a bicontinuous topology similar to NPG ($\tau_{\rm Vycor} = 3.5 \pm 0.3$ \cite{Gruener2011}).

The value of $\tau=3.2$ for NPG confirms the previous findings for contiguity, connectivity and isotropy of the pore space in NPG~\cite{Roesner2007,Fujita2008}. This characteristics is favorable for efficient fluid transport. In order to assess the suitability of the material for capillary pumping, let us inspect the capillary pressure, $P_{\rm c}$ in its pore space. Based on $ P_{\rm c} = 2 \sigma_\mathrm{LV}  \cos \theta / r_0$ ~\cite{Gruener2011}, we obtain the estimated value as  $P_{\rm c} \sim 0.74$  MPa or 7.4 bar. In this context it is also of interest to estimate the average flow rate. For the experiment of Fig.~\ref{OCP} (sample length 1 cm), this parameter is obtained as 21 nL/s$\cdot$mm$^2$. This compares favorably to the result obtained with microstructures of similar length but larger channel size, etched in silicon, which result in 0.7-12nL/s $\cdot$mm$^2$  (ref.~\citenum{Zimmermann2007}).


\noindent{\bf Imbibition at potential.} While the open-circuit experiment is an example for imbibition under charge control, it is also instructive to explore the consequences of controlling the electrode potential on the imbibition kinetics. As a preliminary, we estimate the change of the contact angle with electrode potential and inspect its impact on the imbibition rate using equation (\ref{eqn_LW}).
Changes in contact angle as the function of the electrode potential are routinely studied in electrowetting experiments. As compared to the conventional approach, which uses a planar surface covered with a smooth dielectric layer, electrowetting directly on planar metallic electrodes -- which is analogous to our experiment with a porous metal -- requires a low operating voltage. Such a scheme has been demonstrated for actuating fluidic transportation in microchannels \cite{Satoh2004}.
Electrowetting studies on clean planar gold surfaces, using 1 M KCl aqueous solution, have found a change in  contact angle between 70$^\circ$ at open circuit potential (OCP) to $\sim$30$^\circ$ at -1 V (vs. Ag/AgCl) \cite{Yokomaku2008}. In order to compare the wetting properties of 1M KCL and 1M KOH on gold, we determined the imbibition coefficient for 1M KCL. The result, $C$ = 0.37 mg$\cdot$s$^{1/2}$, is well compatible with the $C$ = 0.36 mg$\cdot$s$^{1/2}$ for  1M KOH as discussed above. This suggests that similar changes in contact angle can be expected for both systems. Therefore, according to equation (\ref{eqn_LW}), an acceleration of imbibition by $\sim$60$\%$  can be expected in potential range between OCP and -1V. We have tested that prediction, as will now be explained.

\begin{figure*}
  \includegraphics[width=18cm]{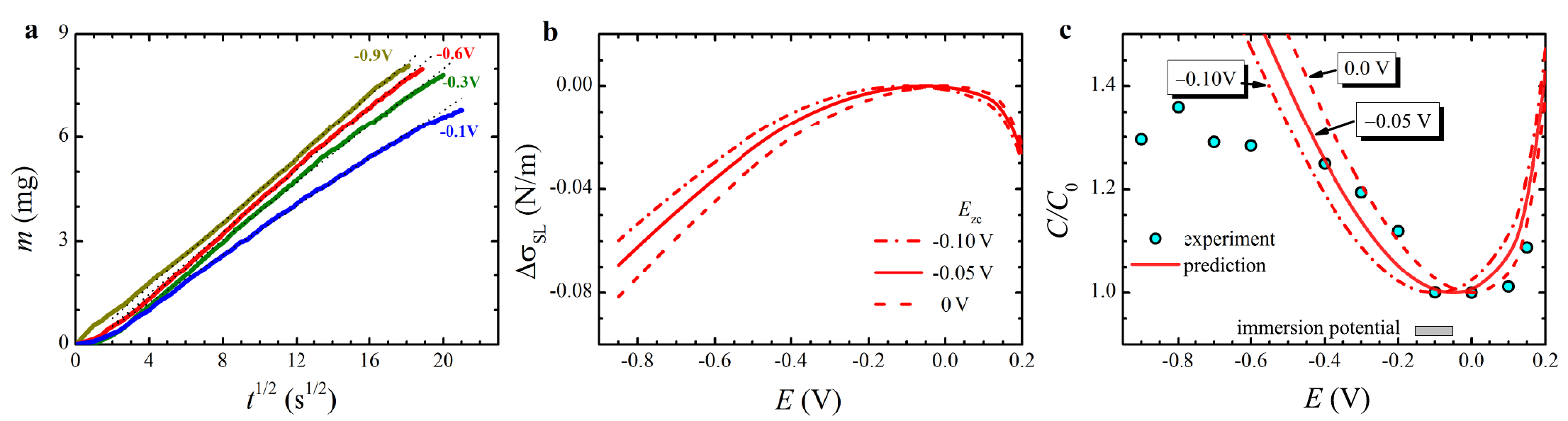}\\
  \caption{{\bf Imbibition kinetics versus air.} \textbf{a}, Imbibed mass, $m$, versus square-root of time, $t$, for experiments at the potentials indicated by the labels. Solid lines: experiment, dashed lines: linear fits of experimental data for $t$ $>$ 25 s. \textbf{b}, Solid-liquid interfacial tension change, $\Delta \sigma_\mathrm{SL}$, versus electrode potential, $E$. Dash-dotted line: the potential of zero charge, $E_\mathrm{zc}$ = -0.10 V, solid line: $E_\mathrm{zc}$ = -0.05 V, dashed line: $E_\mathrm{zc}$ = 0 V. \textbf{c}, Imbibition coefficient, $C$, normalized to the imbibition coefficient, $C_0$, at the open-circuit potential, versus $E$. The dash-dotted, dashed and solid lines interpolating the data points are predictions of $C/C_0$ based on the data of $\Delta \sigma_\mathrm{SL}$ by setting $E_\mathrm{zc}$ = -0.10, -0.05 and 0 V, respectively. Grey bar: the range of immersion potential. }
  \label{fig_imbi_potential}
\end{figure*}

Imbibition of electrolyte by NPG at potential was studied in the potential range between -1 V  and +0.2V. The upper limit was imposed by the need to avoid gas bubbles from oxygen evolution. Mass transients during imbibition at four discrete potential values are shown in Fig. \ref{fig_imbi_potential}a. Linear graphs, similar to those of Fig. \ref{OCP}, confirm that the Lucas-Washburn law holds for the experiments at potential.

\begin{figure*}
  \includegraphics[width=8.5cm]{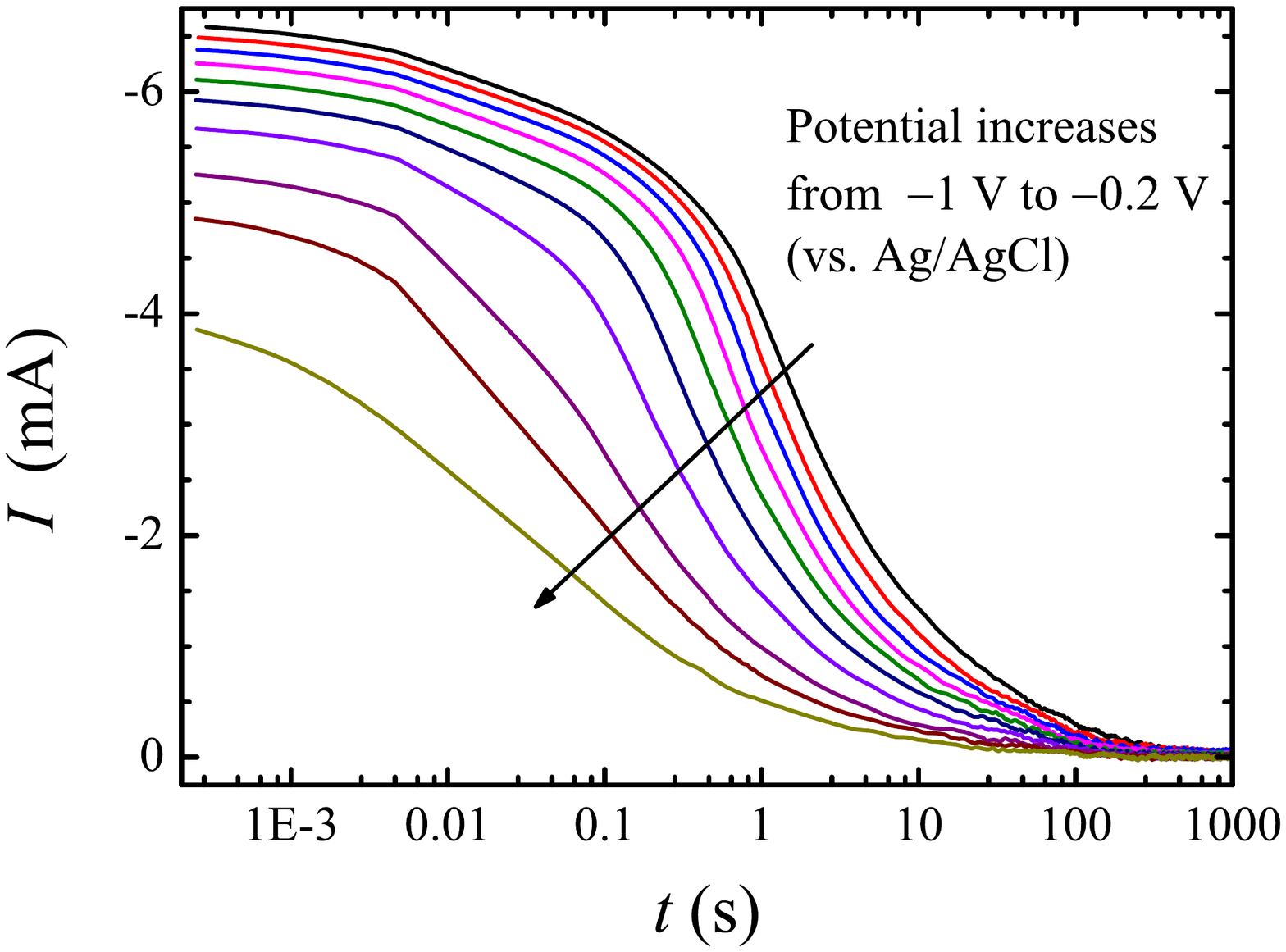}
  \caption{{\bf Charging kinetics.} Step coulometry data on a fully wetted sample. Transients of current, $I$, versus  time, $t$, after stepping from the initial potential of -0.1 V to various final potentials $E_{\rm f}$. Between successive graphs, $E_{\rm f}$ varied from -1 V to -0.2 V with a step of 0.1 V.}
  \label{fig_current}
\end{figure*}

The above linearity indicates that the potential-dependent contact angle remains at equilibrium during imbibition. In other words, the ionic transport is sufficiently fast to ensure superficial charging near equilibrium throughout the wetted pore space and up to the imbibition front.
By means of verification, we investigated the charging kinetics in a fully wetted sample by step coulometry, that is, by recording transients of charging current subsequent to potential jumps. The potential was initially held at a reference value of -0.1 V, close to the `potential of zero charge' (PZC) until no current flowed. The potential was then switched to the value of interest and kept constant while the current transient was recorded. The results, which are plotted in Fig.~\ref{fig_current}, indicate a half-current-time scale of 0.1 and 1s. This is indeed much faster than the imbibition times in all experiments.

Figure~\ref{fig_imbi_potential}c shows the potential dependence of the imbibition coefficients, $C$, which are obtained as the slopes of straight lines of best fit to the data of Fig.~\ref{fig_imbi_potential}a. For display in the figure, the coefficients were normalized to the imbibition coefficient, $C_0$, at the OCP. The value of the latter was -50 mV during the experiments described in the previous section. It is apparent that the experiment puts $C$ at minimum (slowest imbibition) at the OCP. This potential-dependence will now be discussed.

The solid-electrolyte interfacial tension, $\sigma_\mathrm{SL}$, obeys the Lippmann equation \cite{Lippmann1875},
\begin{equation}
\label{Eq:Lippmann_equation}
\centering
\mathrm{d} \sigma
_\mathrm{SL} = - q \mathrm{d} E
\end{equation}
with $q$ the charge density (charge per unit area). As the capacitance is always positive, the surface tension is at maximum at the PZC, where $q=0$. The decrease in $\sigma_\mathrm{SL}$ upon charging will enhance the wetting. This is a well-known consequence of the Young equation, $\sigma_\mathrm{LV} \cos  \theta =\sigma_\mathrm{SV}-\sigma_\mathrm{SL}$, where $\sigma_\mathrm{SV}$ is the solid-vapor interface energy. Thus, a minimum in the imbibition rate is expected at the PZC.

A rough estimate of the PZC can be obtained independently by the immersion method \cite{Bockris1969}, that is, by recording the open-circuit potential when the dry sample is immersed in 1 M KOH. The immersion potential was found at -100 $\pm$ 50 mV. This is indeed close to the potential of minimum $C$.

The expected impact of potential on the imbibition rate may be quantified based on the variation, $\Delta \sigma_\mathrm{SL}$, of the solid-liquid interfacial tension as computed by integrating the Lippmann equation.
To this end a first integration of the experimental CV of Fig.~\ref{fig_setup}c yields $q(E)$, with the constant of integration set by the PZC. The minimum in $C$, as our most reliable relevant observation, puts the PZC at $E_{\rm zc} = -0.05 \pm 0.05 \rm V$.
A second integration then yields $\Delta \sigma_\mathrm{SL}$, as shown in Fig.~\ref{fig_imbi_potential}b.
As a measure for the acceleration of the imbibition, the ratio $C/C_0$ is then obtained via the relation -- readily verified by comparison to equation (\ref{eqn_LW_coefficient}) -- $C/C_0 = (1 - \Delta \sigma_\mathrm{SL} /(\sigma_\mathrm{LV} \cos \theta_0 ) )^{1/2}$.
Figure ~\ref{fig_imbi_potential}c compares the result to the experimental imbibition rate constants. It is seen that the two independent approaches to the imbibition rate agree reasonably well in the vicinity of the PZC. For potentials farther from the PZC, the slower-than-predicted imbibition is compatible with the empirically well established contact-angle saturation in electrowetting\cite{Mugele2005,Kornyshev2010}. These findings support our discussion of electrocapillarity as the origin of the accelerated imbibition.


\noindent{\bf Imbibition into cyclohexane-saturated NPG.} Our results so far demonstrate that electrocapillary effects can be used to accelerate the capillary flow in NPG. In order to realize even stronger control on the imbibition kinetics, we probe the strategy that replaces air as the pore fluid with cyclohexane.
This is mainly motivated by the low interfacial tension of the metal-cyclohexane interface. As a consequence, the driving force for imbibition of the aqueous electrolyte is reduced when the pores are initially wetted by cyclohexane. A small change in the electrolyte-metal surface tension, as can be achieved by varying the electrode potential, may then be sufficient to significantly change or even invert the imbibition.
Cyclohexane has similar viscosity to water. Its lower density and immiscibility with water enable it to float on water. It is known that smooth metal surfaces are generally more oleophilic than hydrophilic \cite{Fox1955}, so that once NPG is saturated with cyclohexane, it is impossible to imbibe it with an aqueous electrolyte. However, electrocapillary effects provide the opportunity to tune the interface energy between the electrolyte and metallic electrode surfaces, and thus control the relative wettability of water and cyclohexane on the metal surface.

\begin{figure*}
  \includegraphics[width=14cm]{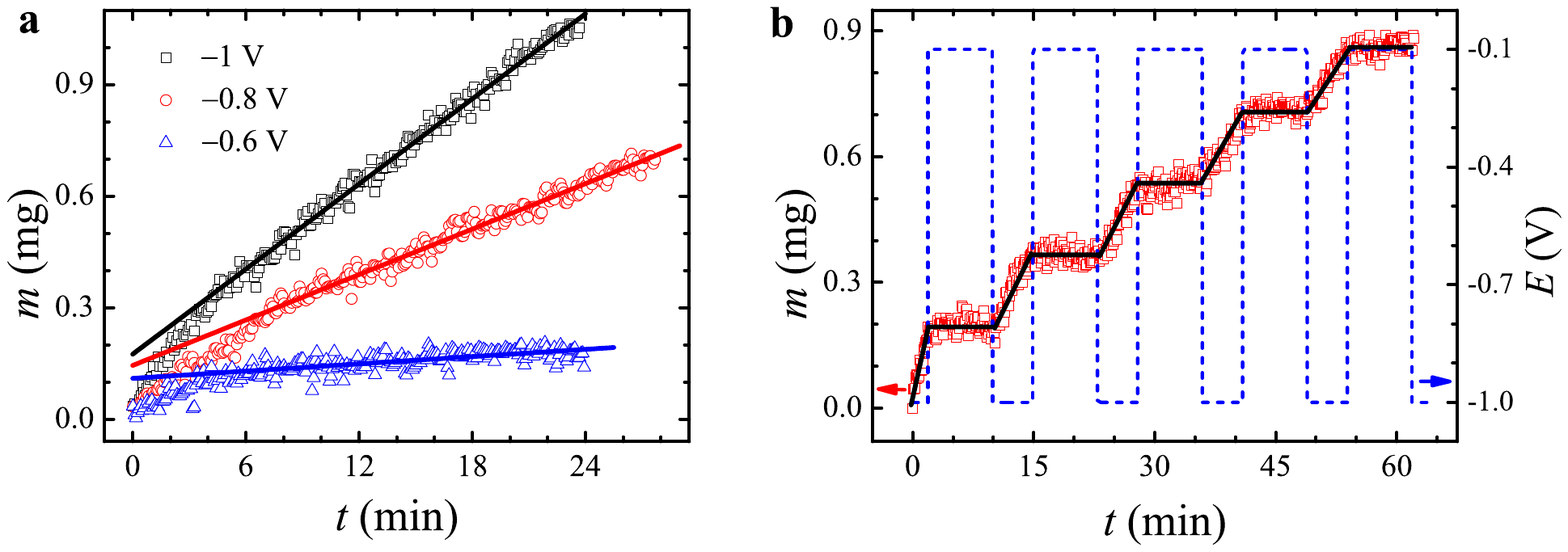}\\
  \caption{{\bf Imbibition kinetics versus cyclohexane.} \textbf{a}, Mass, $m$, versus time, $t$, for experiments at various constant values of the electrode potential, as indicated by labels. Solid lines are straight-lines of best fit. \textbf{b}, Response of imbibition to step potentials. Symbols and left ordinate: mass, $m$, evolution with time, $t$, during imbibition of 1M KOH into  cyclohexane-saturated nanoporous gold. The potential was stepped as indicated by the dashed line and right ordinate. The black solid line is a guide to the eye.}
  \label{fig_replace_V}
\end{figure*}

When cyclohexane-saturated NPG is brought in contact with water, there is indeed no imbibition. The situation changes when a potential is applied.
This is apparent in Fig.~\ref{fig_replace_V}a, which shows mass change versus time for different, constant values of the electrode potential. The monotonic increase of mass with time at each potential confirms that the electric potential makes the gold surface more hydrophilic, which prompts the imbibition. As can be seen in the figure, a more negative potential leads to a higher imbibition rate. It is remarkable that the higher potentials result in a linear variation of mass with time. This is indeed compatible with theory, as will now be explained.

For the two-phase capillary flow in pores, the Lucas-Washburn equation is rewritten as \cite{Alava2004}
\begin{equation}
\centering
\frac {{\rm d} h} {{\rm d} t} = \frac {{r_0}^2}{8 \tau} \ \frac {P_\mathrm{c}}{\eta_\mathrm{w} h +\eta_\mathrm{c}(L-h)}
\label{eqn_2phase}
\end{equation}
where $h$ is the rising height, $h(t) = m(t)/(\rho_\mathrm{w}-\rho_\mathrm{c}) A \phi_0$ and $L$ is the sample length. The subscripts $w$ and $c$ refer to aqueous electrolyte and cyclohexane, respectively. In view of the fact that $\eta_c$ ($\sim$ 0.993 cP) is very close to $\eta_\mathrm{w}$ ($\sim$ 1.054 cP), we may inspect the special case when the two viscosities are equal. Here, equation (\ref{eqn_2phase}) simplifies to
\begin{equation}
\centering
\frac {{\rm d} h} {{\rm d} t} = \frac {{r_0}^2}{8 \tau} \frac { P_\mathrm{c}}{\eta L} \,.
\label{eqn_2phase_equal}
\end{equation}
The time-derivative of $h$ emerges as a constant. In other words, the theory agrees with our experimental observation in showing imbibition proceeding at a constant rate.

A constant and controllable flow rate is greatly desired in capillary pumping systems~\cite{Zimmermann2007}. Not only do the data of Fig.~\ref{fig_replace_V}a exhibit such a linear law, but it also demonstrate that the imbibition rate is controllable through the electrode potential. For example, at -1 V, the imbibition rate is $6.35 \times 10^{-4}$ mg/s or 3.2 nL/s obtained by the linear fit in Fig.~\ref{fig_replace_V}a, which yields a capillary driving pressure of 0.23 MPa according to equation (\ref{eqn_2phase_equal}). Similarly, a much smaller imbibition rate of 0.27 nL/s is achieved when a potential of -0.6 V is applied.


\noindent{\bf Switchable imbibition.} The observation of a strong correlation of the imbibition rate during cyclohexane replacement to the electrode potential raises the question, can the imbibition in this geometry be completely switched on and off by the potential, or can the direction of the imbibition even be reversed? We have investigated this question by repeatedly switching the potential during an individual imbibition run. The results are shown in Fig.~\ref{fig_replace_V}b and will now be discussed.

Figure~\ref{fig_replace_V}b shows an experiment where the sample was first contacted by the electrolyte while a negative potential (-1 V) was applied. The potential was then repeatedly switched between -1V and -0.1 V (near the PZC). It is seen that mass increases rapidly after the initial contact at -1 V, indicating spontaneous imbibition. When the potential is switched to -0.1 V, the mass increase immediately stops, in other words, imbibition is switched off. As can be seen in the figure, this on-off switching sequence is perfectly repeatable over several cycles. In our experiments, this behavior was only limited by the finite capacity of the sample for imbibition of KOH solution.

It would be equally interesting to be able to invert the sequence and, with a suitable potential value, prompt the sample to imbibe cyclohexane, replacing the aqueous electrolyte. However, we found that imbibition of KOH is arrested in the entire potential regime between -0.4 and 0.2 V, which gives the system a valve-like behaviour.

This behaviour could be attributable to a pronounced electrowetting hysteresis \cite{Mugele2005, Kornyshev2010, Li2012}, that is a hysteresis of the contact angle as a function of applied voltage. This, however, is unlikely in view of the reproducible electrowetting behaviour for the aqueous electrolyte (see Fig. \ref{fig_imbi_potential}), which is well describable by classical theory.

The imbibition arrest may also originate in meniscus pinning in the pore network at obstacles such as ligament and pore junctions. This features of the local geometry can lead to advancing or receding interfaces with vanishing mean curvature, and thus vanishing Laplace pressures. This phenomenon, which is particularly likely for contact angles greater than a certain threshold angle $\theta_\mathrm{tr} < 90$ deg \cite{Concus1969, Mognetti2010, Chibbaro2009, Paxson2013}, is known as Gibbs pinning \cite{Gibbs1969}. We calculated the Laplace pressures (and therefrom the mean contact angle) with equation (\ref{eqn_2phase_equal}) from our imbibition experiments to $\theta$=71$\pm2$, 81 $\pm2$  and 84 $\pm2$ degree for -1, -0.8, and -0.6 V, respectively. We observe unpinning of the interface at -0.6 V during voltage lowering, which corresponds to a $\theta_\mathrm{tr}= 84\pm2$ degree. As this value is close to 90 deg, it indeed suggests at least weak Gibbs pinning. Such a pure geometrical arrest mechanism is also plausible if one takes a simplified view and considers the 3-D isotropic pore space as representable by three sets of parallel-aligned arrays of ligaments in the three spatial directions. In the case of flow along one ligament direction, the two sets of perpendicularly aligned ligament arrays act like posts in the flow path, which according to recent mesoscale simulations for electrowetting in microchannels can indeed result in Gibbs pinning \cite{Mognetti2010}.

Note, however, that a pure wetting angle hysteresis between advancing and receding liquid/liquid interface, as it can be induced by roughness on macroscopically planar surfaces \cite{deGennes2004, Marmur2006}, could also translate in imbibition hysteresis in solids with rough porous surfaces \cite{Bear1972, Murison2013, Raeesi2013}.
In view of the crystalline nature of NPG, its highly curved surfaces inherently require an abundance of step edges and kinks. In other words, the nanoscale curvature presupposes atomic-scale roughness. Yet, this observation may also be viewed as suggesting that the small ligament size forbids a meaningful separation between intrinsic microstructural features (such as step edges) and extrinsic surface roughness. This distinguishes nanoporous materials from planar surfaces or simple channel geometries, rendering a direct comparison challenging, if not impossible.

Simulation studies on simplified NPG representations and pore networks \cite{Sadjadi2013} along with studies on the evolution of the imbibition front width as a function of time \cite{Gruener2012} and complementary examinations of the imbibition front by electron microscopy \cite{Paxson2013} may allow for more detailed insights with regard to the distinct pinning mechanisms and a relative quantification of their contribution to the wetting hysteresis documented here.

We can imagine that imbibition reversal is possible, if cyclohexane is replaced with a fluid with a smaller interfacial tension at its interface with gold. Such a fluid would have a higher driving force for replacing the aqueous phase. As an alternative approach, the cyclohexane -- or any other second fluid -- could be pressurized, again increasing the propensity for reverse imbibition.

\noindent{\bf Discussion}

Whereas studies of imbibition in the past used nonmetallic host structures, we have here explored a metal with a nanoscale pore structure as the imbibition host. We find that the spontaneous imbibition of aqueous electrolyte, when displacing air, obeys the Lucas-Washburn law. A value of tortuosity 3.2 $\pm$ 0.2 for NPG reveals a good connectivity and isotropic property of the pore structure, which favors fluid flow.
As the host structure is electrically conductive, we can control the electrode potential and, thereby, tune the solid-electrolyte interfacial tension. This allows us to control the driving force for imbibition, and the results show that the imbibition kinetics can be significantly accelerated. By replacing air with cylcohexane as the second pore fluid, the competition between wetting by cyclohexane and aqueous electrolyte makes the surface effectively less hydrophilic. This suggests that small changes in the metal-electrolyte interfacial tension, as they can be induced by changes in the electrode potential, may be sufficient to switch between hydrophobic and hydrophilic behavior in respect to the aqueous electrolyte.
Indeed, we could show that spontaneous imbibition into cyclohexane-saturated nanoporous gold is suppressed on the charge-neutral surface. Transferring electronic charge onto the surface by applying a potential prompts the imbibition to start. In other words, we have demonstrated that imbibition in a nanoporous metal can be switched on or off by electric signals.
Our results demonstrate that nanoporous gold as an imbibition host structure provides high flow rate and large capillary pressure. Together with the demonstration of switching, the findings suggest that nanoporous gold may be used as an accurately controllable capillary pump.
In that respect it is noteworthy that the pore size and porosity of NPG can be controlled by using different synthesis procedures or thermal treatment~\cite{LiSieradzki1992, Wang2013,Kertis2010}. In this way, the flow rate and capillary pressure in NPG can be tailored. Furthermore, NPG is readily made in different shapes and dimensions. Reports range from nanoporous nanoparticles \cite{WangD2012}) and submicron porous membranes \cite{Ding2004} over micron-sized pillars \cite{Volkert2006}) to wires (as in the present work) or plates of mm to cm dimensions \cite{Wang2013}. While our samples were fabricated based on metallurgical casting and cold working, nanoporous gold has also been made from vapor- or electrodeposited alloy layers~\cite{Lee2007}.
This process is readily integrated into the lithographic process technology that goes into the fabrication of devices for micro-electro-mechanical or nanofluidics applications. Our findings therefore suggest that nanoporous metal may be implemented in devices as a way to manipulate femtolitre to microlitres amounts of liquid under control of an electric potential.



\noindent{\bf Methods}

\noindent{\bf Fabrication of nanoporous gold.} Samples were prepared by electrochemical dealloying of arc-melted ingots of Ag$_{75}$Au$_{25}$ after homogenization, wire-drawing and cutting into size (i.e., 1.08 $\pm$ 0.01  mm in diameter and 10 $\pm$ 0.5 mm  in length). Electrochemical annealing by cyclic scanning over a wide range of electrode potential lead to structural coarsening, resulting in a uniform ligament size of 45 nm, removed adsorbed oxygen species and residual Ag. For details of the preparation see ref.~\citenum{Jin2011}. The samples were then repeatedly rinsed in ultra-pure water, soaked in pure alcohol ($\geq$ 99.9\%, Merck KGaA Frankfurt) for 12 hours and dried in pure Argon (99.5\%) flow for over 2 days. For capillary flow measurements, the samples were covered with solvent-resistant epoxy adhesive (Loctite 3450, Loctite Germany GmbH, Munich, Germany) by painting but left uncovered at their top and bottom ends.

\noindent{\bf Imbibition versus air.} The electrolyte for imbibition experiments was 1 M KOH (aq) (analytically pure, Merck KGaA Frankfurt). The schematic of the experimental setup is shown in Fig.~\ref{fig_setup}a. The electrochemical characterization and control used a a potentiostate (Autolab M101, EcoChemie) and a three electrode system with the NPG sample as working electrode (WE), a commercial Ag/AgCl in 3M KCl reference electrode (RE) and NPG with larger surface area as the counter electrode (CE). The RE is put inside a Luggin capillary, the  other end of which approaches the NPG sample closely  during under-potential measurements. The total surface area was determined from the CV, exploiting the known value, $50 \rm \mu C/m^2$, of the charge transferred to the electrode in the first 150 mV of potential after the onset of OH-adsorption \cite{Lipkowski1999}. For the imbibition studies, the sample is suspended from a microbalance while contacted by a gold wire. The electrolyte surface in the reservoir is raised at a velocity of 10 $\mu$m/s by a motorized laboratory syringe pump until the electrolyte contacts the sample. The sample mass evolution during the imbibition is sampled at the rate 1 s$^{-1}$.

\noindent{\bf Step coulometry method.} Step coulometry was carried out by partially immersing the adhesive-covered NPG sample in the electrolyte (i.e. 1 M KOH) and monitoring current transients in response to potential steps after the sample was full of the electrolyte. A reference potential of -0.1 V was chosen, which is close to the potential of zero charge. Before switching to a new potential, the sample was first equilibrated at -0.1 V, and then kept at the new potential until the current decayed to almost 0.

\noindent{\bf Imbibition versus cyclohexane}. For the experiment of aqueous electrolyte imbibition into cylcohexane-preimbibed NPG, the experiment setup is the same as shown in Fig.~\ref{fig_setup}a, except that the aqueous solution was covered with a thick layer of cyclohexane ($\geq$ 99.5\%, Sigma-Aldrich). The sample was first allowed to imbibe cyclohexane by partial immersion into that layer, and then the electrolyte column was raised up by gradually adding electrolyte until the sample was touched by the cyclohexane/electrolyte interface. For both constant and step potential measurements, to insure the adequate wetting of the lower sample end by electrolyte, a potential of -1 V was applied for 2 min, and then, the potential was kept at -0.1 V for 8 min to stabilize the outside meniscus. During different potential were applied, the replacement of cyclohexane by the electrolyte was monitored from the balance reading.

\noindent{\bf Acknowledgement} This study was funded by the Helmholtz-Chinese Scholarship Council fellowship program (Y. Xue) and the Alexander von Humboldt Foundation Group Linkage Program (H. Duan and J. Weissm\"uller). Y. Xue acknowledges support of his PhD studies by the National Natural Science Foundation of China grant 10932001. The work also profited from the German Research Foundation (DFG) through SFB 986 ''Tailor-Made Multi-Scale Materials Systems  M3'' and the DFG priority program SPP1164 ''Nano- and Microfluidics'' grant No. Hu850/2.

\noindent{\bf Author contributions}
Y.X., J.W. and H.D. conceived the experiments. Y.X., J.M. and P.H. contributed to design the experiments.  Y.X. carried out the fabrication of the samples and performed the experiments.  Y.X., J.W. and P.H. wrote the paper, and all authors discussed the results and commented on the manuscript.

\noindent{\bf Additional information}
The authors declare no competing financial interests. Reprints and permission information is available online at http://npg.nature.com/reprintsandpermissions/. Correspondence and requests for materials should be addressed to P.H.

\end{document}